# Multi-Party Private Set Intersection: A Circuit-Based Protocol with Jaccard Similarity for Secure and Efficient Anomaly Detection in Network Traffic

Jiuheng Su

Shanghai Key Laboratory of Trustworthy Computing, East China Normal University, 51215902053@stu.ecnu.edu.cn

Zhili Chen*

Shanghai Key Laboratory of Trustworthy Computing, East China Normal University, zhlchen@sei.ecnu.edu.cn

Xiaomin Yang*

Xinhua Hospital, School of Medicine, Shanghai Jiao Tong University, yangxiaomin@xinhuamed.com.cn

We present a new circuit-based protocol for multi-party private set intersection (PSI) that allows $m$ parties to compute the intersection of their datasets without revealing any additional information about the items outside the intersection. Building upon the two-party Sort-Compare-Shuffle (SCS) protocol, we seamlessly extend it to a multi-party setting. Demonstrating its practicality through implementation, our protocol exhibits acceptable performance. Specifically, with 7 parties, each possessing a set size of $2^{12}$, our protocol completes in just 19 seconds. Moreover, circuit-based protocols like ours have an advantage over using custom protocols to perform more complex computation. We substantiate this advantage by incorporating a module for calculating the Jaccard similarity metric of the private sets which can be used in the application domain of network traffic analysis for anomaly detection. This extension showcases the versatility of our protocol beyond set intersection computations, demonstrating its efficacy in preserving privacy while efficiently identifying abnormal patterns in network flow.

## 1 INTRODUCTION

In the problem of private set intersection (PSI), multiple parties possess individual sets of items and aim to ascertain only the intersection of these sets, without revealing any additional information. Currently, whether it's a custom PSI protocol or a generic PSI protocol, two-party PSI is a practical cryptographic primitive with extremely fast and secure implementations [1, 2]. It could even be argued that, to some extent, these two-party PSI protocols only slightly lag behind the most naïve and non-secure method of exchanging hashed values directly in terms of performance. In fact, the problem of PSI has always been a hot issue in the field of secure multi-party computation (MPC). PSI (both two-party and multi-party) has many privacy-preserving applications such as private contact

*Zhili Chen is the corresponding author, and Xiaomin Yang is the co-corresponding author.
This work is supported by the Natural Science Foundation of Shanghai (Grant No. 22ZR1419100), the National Natural Science Foundation of China Key Program (Grant No. 62132005), and CAAI-Huawei MindSpore Open Fund (Grant No. CAAIXSJLJJ-2022-005A).

discovery [3], private healthcare information sharing and so on [4]. Many companies, such as Google and Facebook, employ PSI for mining and utilizing shared information. With the increasing prevalence and scale of private data sharing, there is a growing emphasis on the sharing and utilization of private data. The multi-party PSI problem will also gain broader recognition and research attention due to its more extensive applicability.

In our work, we focus on the multi-party PSI in the semi-honest model, which is a natural generalization of the two-party PSI. By "semi-honest" we refer to threat model where all parties will follow the protocol, but adversaries may attempt to extract as much information as possible from the protocol execution. We have observed that, as commonly acknowledged, achieving secure computation in a multi-party setting poses significant challenges, often accompanied by complex communication interactions. Existing protocols in generic MPC, such as garbled circuits, tend to become more intricate and costly when extended to multi-party setting. However, our protocol, serving as the circuit-based implementation for multi-party PSI, mitigates the complexity associated with multi-party interactions, demonstrating simple scalability and reasonable computation and communication overhead.

### 1.1 State of the Art for Multi-party PSI

The first multi-party PSI was proposed by Freedman, Nissim and Pinkas [5], which utilized oblivious polynomial evaluation (OPE) techniques like additively homomorphic encryption. The recent work by Ghosh and Nilges [6] replaces the expensive homomorphic encryption with an efficient protocol for oblivious polynomial evaluation (OLE). The first practical multi-party PSI protocols was proposed by Kolesnikov et al. in [7], which is implemented by oblivious transfer (OT) extension [8] and oblivious programmable pseudorandom function (OPPRF) they introduced. This protocol is designed with two variants, one tailored to resist semi-honest adversaries and the other specifically crafted to thwart augmented semi-honest adversaries. Another independent effort conducted by Ben Efraim et al [9]. They introduce the initial concretely efficient maliciously secure multi-party PSI. This work ingeniously integrates findings from semi-honest multi-party PSI and malicious two-party PSI, both relying on the utilization of garbled bloom filter (GBF).

To the best of our knowledge, the research attention on circuit-based multi-party PSI is limited in current literature, likely due to the perceived high computational and communicational costs associated with generic secure multiparty computation protocol. However, our results demonstrate that a careful implementation of garbled circuits yields solutions capable of operating practically even on million-element sets.

### 1.2 Overview of our Protocol

Our protocols are constructed upon the groundwork laid by two-party circuit-based PSI protocol proposed by Huang [10], we diverge from the focus on the BWA protocol, which is limited to small data domains. Instead, our emphasis is directed towards the two-party Sort-Compare-Shuffle (SCS) protocol. Our protocol also follows the SCS paradigm and extends it to multi-party setting. Hence, we refer to it as Ex-SCS. A distinctive feature lies in our approach to address the intricate challenges of interactions and scalability in the multi-party context, which is secure against an arbitrary number of colluding semi-honest parties.

Before conducting the circuit computation, the private inputs of the involved parties must be securely shared between the two parties (usually $P_1$ and $P_2$) participating in the secure computation process. Subsequently, these two parties collaborate to reconstruct private items and perform secure SCS computations for multi-party PSI within the circuit. Specifically, $P_1$ and $P_2$ sequentially execute the two-party SCS protocol using the ordered data from each party, with $P_1$ inputting the sorted version of its sequence or the resulted sequence of the previous execution, and $P_2$ inputting one sequence from the remaining. If we assume that the number of participants is $m$, we can see that this extension requires $m-1$ calls of two-party SCS protocol. Despite the potential presence of some constant factors in scenarios with a larger number of parties, our protocol overall still maintains the $O(nlogn)$ complexity of the two-party SCS protocol for achieving the computation through comparisons. Additionally, to highlight the scalability advantage of our protocol, we have incorporated counters within the circuit to calculate the count of elements in the intersection for Jaccard similarity computation. This integration can be applied in the practical application of secure anomaly detection in network traffic across different nodes.

## 2 PRELIMINARIES

### 2.1 Setting

We use the following notation in our protocol. There exist $m$ parties, denoted as $P_1, ..., P_m$ where $P_1$ and $P_2$ typically serve as the primary parties for secure computation. Each of these entities possesses specific input sets, namely $S_1, ..., S_m$, with each set containing $n$ elements represented by $\sigma$ bits. And we use $S_i[j]$ to denote the $j$-th item in the set $S_i$. We also denote the computational and statistical security parameters by $\kappa$ and $\lambda$. We use $Q(S)$ to denote the ordered version of the sequence $S$, with the default assumption of ascending order.

### 2.2 Secure Computation

In current research on MPC, two prominent methods for achieving generic secure computation are GC protocol [11] and the GMW protocol [12]. GC protocol is characterized by a constant round complexity and incorporates the Free XOR gates technique [13]. Through half-gate and optimization techniques outlined in [14], the protocol requires two or even fewer ciphertext transmissions to evaluate an AND gate. Similarly, the GMW protocol also utilizes the Free XOR technique and involves two ciphertext transmissions to evaluate each AND gate by OT extension. Many PSI applications do not require the intersection itself, but rather specific functional computations over the items in the intersection. In this context, the advantage of generic protocols becomes apparent, as they offer the flexibility to extend the functionality of the protocol without compromising security.

### 2.3 Two-party Sort-Compare-Shuffle Protocol

A basic PSI circuit computes $O(n^2)$ element comparisons, resulting in $O(\sigma n^2)$ gates, where $\sigma$ represents the bit-length of the elements. However, the two-party SCS protocol, following the paradigm of Sort-Compare-Shuffle, fully utilizes the local computing capabilities of the participating parties. By having the parties perform local sorting in advance, it reduces the overall number of comparisons to $O(nlogn)$.

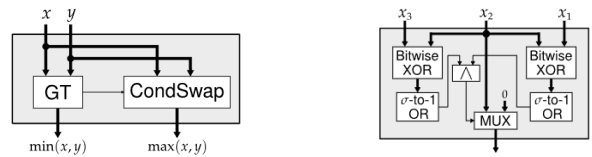

Figure 1: The foundational modules of the two-party SCS Protocol: 2-Sorter (left) and 3-DupSelection (right) circuit.

The protocol can be divided into three main steps: sorting, comparing, and shuffling. Once the local sorting is completed, the private sets of both parties are fed into the circuit for secure computation. By utilizing a bitonic sorting network, which employs basic comparison and exchange operations, the sets are merged into a single, sorted sequence. The term "bitonic" refers to a sequence that is constructed by concatenating an ascendingly sorted sequence with a descendingly sorted sequence. This is the reason why local sorting is necessary prior to the secure computation. The entire sorting process entails $O(nlogn)$ comparisons and exchanges, which are executed by the fundamental module 2-Sorter circuits (see Figure 1). The overall cost of the entire protocol predominantly resides in this initial part. After successfully performing the merge sort on $2n$ elements, it is guaranteed that any elements in the intersection will be adjacent. The determination of intersection elements can be achieved by comparing adjacent positions. The elements in the intersection are matched by a sequence of Duplicate-Selection circuits (see Figure 1) that incorporate built-in multiplexers. When a match is successful, the element is output. In case of a failed match, an explicitly defined invalid dummy value $0^\sigma$ is output to preserve the data obliviousness property of the garbled circuit. Because the process of matching the intersection elements involves linear comparisons, it can be concluded that only $O(n)$ equality comparisons are executed. In addition, by capitalizing on the property that the initial set does not consist of duplicate elements, the overhead can be further minimized through the utilization of 3-DupSelection, simplifying the circuit design. After filtering out all the elements from the intersection, only dummy values and the elements from the intersection are left. However, due to the previous sorting operations, the relative positions of the elements in the intersection and the dummy values may still unintentionally reveal information about the initial sets of the parties. Hence, at this point, the current results cannot be openly disclosed. For instance, if the last element is a non-dummy value $y$, it can be deduced that the initial sets of both participating parties surely include the maximum element $y$. In order to disrupt this positional relationship, three different methods were employed in [10]. However, in order to uphold the consistency of the GC protocol, this paper exclusively employs the Shuffling network to shuffle the elements prior to publicly disclosing the intersection results. This step requires only $O(nlogn)$ symmetric-key operations, making it the least computationally expensive among the three stages.

## 3  CIRCUIT-BASED MULTI-PARTY PSI PROTOCOL

In this section, we provide a detailed exposition of the proposed circuit-based multi-party PSI protocol. As shown in Figure 2, our multi-party PSI protocol proceeds in two consecutive phases: secret set sharing and reconstruction, followed by Ex-SCS secure computation.

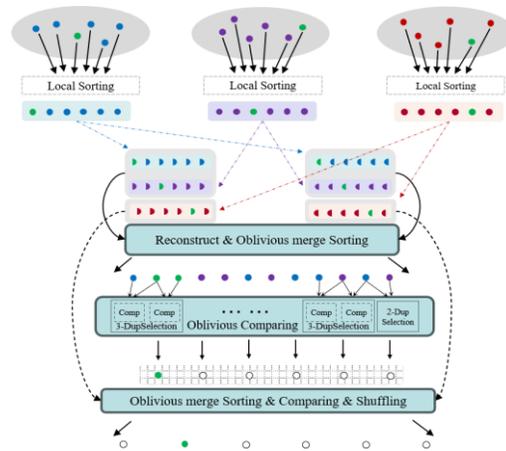

Figure 2: Multi-party Ex-SCS protocol to compute the intersection $I = SCS(SCP(S_1, S_2), S_3)$. "*SCP*" means Sort-Compare-Permute.

## 3.1 Secret Set Sharing and Reconstruction

As shown in Figure 3, we use a simple additive secret sharing scheme to accomplish secret set sharing and reconstruction. The whole reconstruction process of the circuit is relatively clear and the circuit can be obtained by instantiating a binary XOR gate $mn\sigma$ times. The use of the free-XOR gates technique enables the evaluation of XOR gates in the circuit without incurring any communication or cryptographic operations.

> **Parameters**: $m$ parties $P_1, ..., P_m$, where $P_1$ and $P_2$ are the two sides for computing the Boolean circuit $C$.
> **Preparations**: Each party $P_i$ performs local sorting to get the ordered sequence $Q(S_i)$ and picks random seeds $r_{i,j}$ for $j = 1, ..., n$.
> **Execution**:
> 1. Each party $P_i$ sends $\{r_{i,j}\}$ to $P_1$ and $\{r'_{i,j} = r_{i,j} \oplus Q(S_i)[j]\}$ to $P_2$.
> 2. $P_1$ and $P_2$ input the received secrete share $\{r_{i,j}\}$ and $\{r'_{i,j}\}$ and compute $\{r_{i,j} \oplus r'_{i,j}\}$ to reconstruct the sorted sets $Q(S_i)$ for $i = 1, ..., m$ within the circuit.

Figure 3: Secret Set Sharing and Reconstruction.

## 3.2 Extension of Sort-Compare-Shuffle

As depicted in Figure 2, our Ex-SCS protocol is a natural extension of the two-party SCS protocol, adhering to the SCS paradigm. Following the reconstruction of sets from each party in the circuit, $P_1$ and $P_2$ sequentially execute the two-party SCS protocol using the ordered data from each party, with $P_1$ inputting the sorted version of its sequence or the resulted sequence of the previous execution, and $P_2$ inputting one sequence from the remaining. We note that it is imperative to maintain that the number of elements in intermediate results remains equal to $n$ as the maximum possible size of the intersection is $n$. Through ingenious circuit design, we have minimized the complexity of the circuit while ensuring the correctness of the protocol. During the oblivious comparing phase illustrated in Figure 4, we observe that through the utilization of MUX multiplexers and Duplicate -Selection circuit, we can ensure that the number of elements in the intermediate result sets remains equal to $n$ (each Duplicate-Selection circuit outputs one element, and there are totally $n$ Duplicate-Selection circuits). These elements either signify successful matches or are replaced with dummy values $0^\sigma$ in the case of matching failures. We can also leverage this characteristic to validate the correctness of the protocol, as the elements in the final intersection of multiple parties must exist within the intermediate result sets.

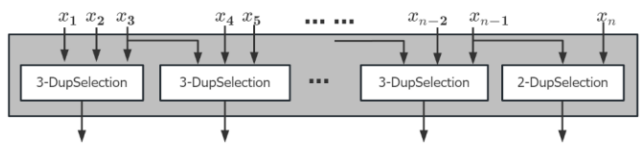

Figure 4: Oblivious comparing circuit design, ensuring that the intermediate results consist of $n$ elements.

Moreover, it is noteworthy that our protocol introduces the shuffling stage only in the last iteration when elements of the intersection are intended to be publicly disclosed. As these are intermediate results, shuffling is unnecessary to conceal the positional information of elements. However, due to the generation of a series of dummy values during the comparison stage for filtering the intersection elements, although the values of the intersection elements are relatively ordered, the overall order of the intermediate results cannot be guaranteed. Here we use a permutation network which can be constructed using $O(n\log n)$ gates to swap all the intersection elements to the end of the sequence. A permutation network can be regarded as a pre-defined circuit that operates on $n$ inputs with a set of control bits. The control bits determine whether specific pairs of elements should be swapped or remain unchanged. By configuring the control bits accordingly, any desired permutation of the $n$ inputs can be achieved. The fundamental component in the permutation network is a Cond-Swap circuit (see Figure 5) that takes two 1-bit values $x, y$ and a control bit $s$ as input. If $s = 0$, the circuit outputs $x$ and $y$ in original order. If $s = 1$, the circuit outputs $x$ and $y$ in swapped order. A 2-Swapper, as required, can be constructed using $\sigma$ Cond-Swap circuits. For our requirement, we can achieve the desired functionality by setting the control bits based on whether the elements in the sequence are dummy values. Specifically, each element in the intersection needs to be swapped with the farthest dummy value to its right, following a descending order.

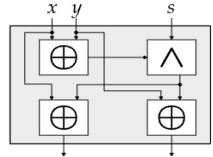

Figure 5: 1-bit Cond-Swap circuit.

## 4 SECURE ANOMALY DETECTION IN NETWORK TRAFFIC

In this section, we elaborate on the application of our multi-party PSI protocol as a submodule in network anomaly traffic monitoring. Traditional network anomaly traffic monitoring relies on rule-based methods such as frequency restrictions, abnormal packet size detection, IP blacklist (known malicious entities) and so on. However, rule-based methods for network anomaly traffic monitoring are limited to single network nodes. If we intend to enhance mutual security by leveraging the anomalous traffic monitoring rules of other network nodes, direct sharing of such rules is evidently impractical, as it would disclose the criteria used by each network node to discern anomalous traffic. In the presence of adversaries, this could potentially facilitate adversaries to circumvent these rules.

Therefore, we propose the utilization of Jaccard similarity to achieve this objective, wherein the computation of Jaccard similarity between one's own new data packets and the anomalous traffic sets of other network nodes serves as an assessment of anomalous behavior. For instance, we employ the widely adopted rule based on the blacklist of IP addresses (if multiple features are considered, they can be combined into feature vectors). The IP addresses from a segment of network traffic over a period are aggregated into a set. Other nodes contribute their respective sets of IP addresses from their blacklists. If the Jaccard similarity between the new data packets' source IP addresses and the anomalous traffic set of a certain node exceeds the predefined threshold $t$, it may indicate that the new data packets exhibit abnormal characteristics and can be labeled as anomalous traffic, indicating a potential anomaly risk in the network; conversely, if the similarity does not meet the threshold, the data packets are categorized as regular. It is noteworthy that a higher threshold may result in a more conservative anomaly detection, reducing false positives but potentially increasing false negatives. Conversely, a lower threshold may enhance sensitivity but might lead to a higher rate of false positives. This depends on the specific application scenario.

Moreover, the generic secure multi-party computation protocols we employ can be easily extended to leverage their intersections for computing Jaccard similarity. By summing the input signals to the MUX inside the Duplicate-Selection circuits (see Figure 1) using a Counter circuit, we can obtain the cardinality of the intersection. For optimization purposes, the counter circuit incrementally increases the number of bits employed to represent its internal state in a deferred manner. With this construction, the counter circuit utilizes $n\log n - n$ non-free gates.

Hence, the utilization of our circuit-based protocol enables the convenient computation of similarity metrics for any number of sets. Our experiment result also indicates that this stage only incurs negligible additional overhead.

## 5 IMPLEMENTATION AND PERFORMANCES

To assess the efficiency of our multi-party PSI protocol, we employed two standard desktop computers featuring 12th Gen Intel(R) Core(TM) i5-12400 2.50GHz processors and 16GB RAM. These computers were connected via a local area network (LAN) with a band-width capacity of 100 Mbps. All experiments were carried out on these computers under the same configurations. We set $\kappa = 128$ for computational security and $\lambda = 80$ for statistical security. Random elements were generated from a predefined universe to ensure uniqueness in each party's set.

### 5.1 Comparison with Custom Multi-party PSI Protocols

The outcomes of an experimental evaluation, aimed at comparing the computational and communicational performance of the Ex-SCS and KMPRT protocol [7] in computing the intersection of extensive input sets, are summarized in table 1. We find that our protocol incurs a concrete communication cost that is 100-1000 times higher compared to a custom multi-party PSI protocol like KMPRT, which discloses the intersection in plain view to the parties. Despite the augmented communication overhead, the obtained results underscore the viability of deploying our Ex-SCS protocols for PSI in large-scale scenarios. This also indicates the potential applicability of our protocol in non-real-time applications where PSI is employed as a submodule, as discussed in Section 4.

Table 1: Running time in second and communication overhead in MB of our multi-party Ex-SCS protocol and comparison with previous work KMPRT in semi-honest setting. All the elements were represented by 32 bits (i.e., $\sigma = 32$).

| $m$ | $n$ | Running Time | | Communication Overhead | |
|---|---|---|---|---|---|
| | | Ex-SCS | KMPRT | Ex-SCS | KMPRT |
| 3 | $2^8$ | 0.48 | 0.16 | 20.19 | 0.61 |
| | $2^{12}$ | 5.29 | 1.22 | 408.46 | 1.14 |
| | $2^{16}$ | 102.93 | 7.64 | 6832.88 | 19.68 |
| 5 | $2^8$ | 0.81 | 0.26 | 68.77 | 0.86 |
| | $2^{12}$ | 12.32 | 2.39 | 1432.14 | 1.58 |
| | $2^{16}$ | 230.17 | 12.38 | 32977.2 | 27.8 |
| 7 | $2^8$ | 1.33 | 0.39 | 160.18 | 1.34 |
| | $2^{12}$ | 19.36 | 3.48 | 2801.66 | 2.42 |
| | $2^{16}$ | 377.41 | 18.56 | 82886.3 | 38.92 |
| 9 | $2^8$ | 1.93 | 0.48 | 237.32 | 1.64 |
| | $2^{12}$ | 26.43 | 4.2 | 3720.98 | 3.24 |
| | $2^{16}$ | 514.65 | 27.26 | 126341.9 | 58.04 |

### 5.2 Extended Evaluation of Our Protocol

To understand the scalability of our protocol in large-scale data scenarios, we evaluate it on the range of the number $n$ in $\{2^{12}, 2^{12}, 2^{20}\}$ and $m = 3$. In the case of $\sigma = \infty$, we considered the same approach as in [15]. Each element is initially hashed to a string of length $40 + 2\,logn - 1$, ensuring collision probability is below $2^{-40}$. Table 2 shows the number of non-free gates required to compute. This metric stands out as a crucial factor influencing the performance of circuit-based MPC protocols. However, in contrast to benchmarks related to communication or runtime, it remains indifferent to the specific details of the MPC implementation. For some secure circuit evaluation protocols like GMW [12] the round complexity depends on the depth of the circuit. Our data is shown in Table 2.

Table 2: Number of non-free gates for each element and the depth of the circuit in our multi-party PSI protocol.

| $n$ | $\sigma$ | Number of non-free gates for each element | Circuit depth |
|---|---|---|---|
| $2^{12}$ | 32 | 1826 | 120 |
| $2^{16}$ | 32 | 2628 | 160 |
| $2^{20}$ | 32 | 3946 | 200 |
| $2^{12}$ | $\infty$ | 2175 | 143 |
| $2^{16}$ | $\infty$ | 3235 | 197 |
| $2^{20}$ | $\infty$ | 4972 | 252 |

## 6 CONCLUSION

The research findings demonstrate that PSI serves as a valuable component in various privacy-preserving applications. Our study reveals that protocols built upon generic secure computation offer the convenience of seamlessly integrating universal secure computations with subsequent computations. Since our multi-party PSI protocols are built using generic garbled circuits, they effectively maintain privacy while efficiently detecting abnormal patterns in network flow. This work provides compelling evidence that numerous secure computation problems can be effectively addressed with the generic secure computation protocol.

**REFERENCES**

[1] Pinkas B, Schneider T, Tkachenko O, et al. Efficient circuit-based PSI with linear communication[C]//Advances in Cryptology–EUROCRYPT 2019: 38th Annual International Conference on the Theory and Applications of Cryptographic Techniques, Darmstadt, Germany, May 19–23, 2019, Proceedings, Part III 38. Springer International Publishing, 2019: 122-153.

[2] Chase M, Miao P. Private set intersection in the internet setting from lightweight oblivious PRF[C]//Advances in Cryptology–CRYPTO 2020: 40th Annual International Cryptology Conference, CRYPTO 2020, Santa Barbara, CA, USA, August 17–21, 2020, Proceedings, Part III 40. Springer International Publishing, 2020: 34-63.

[3] Hagen C, Weinert C, Sendner C, et al. All the numbers are US: Large-scale abuse of contact discovery in mobile messengers[J]. Cryptology ePrint Archive, 2020.

[4] Nevo O, Trieu N, Yanai A. Simple, fast malicious multiparty private set intersection[C]//Proceedings of the 2021 ACM SIGSAC Conference on Computer and Communications Security. 2021: 1151-1165.

[5] Freedman M J, Nissim K, Pinkas B. Efficient private matching and set intersection[C]//International conference on the theory and applications of cryptographic techniques. Berlin, Heidelberg: Springer Berlin Heidelberg, 2004: 1-19.

[6] Ghosh S, Nilges T. An algebraic approach to maliciously secure private set intersection[C]//Annual international conference on the theory and applications of cryptographic techniques. Cham: Springer International Publishing, 2019: 154-185.

[7] Kolesnikov V, Matania N, Pinkas B, et al. Practical multi-party private set intersection from symmetric-key techniques[C]//Proceedings of the 2017 ACM SIGSAC Conference on Computer and Communications Security. 2017: 1257-1272.

[8] Yadav V K, Andola N, Verma S, et al. A survey of oblivious transfer protocol[J]. ACM Computing Surveys (CSUR), 2022, 54(10s): 1-37.

[9] Ben-Efraim A, Nissenbaum O, Omri E, et al. Psimple: Practical multiparty maliciously-secure private set intersection[C]//Proceedings of the 2022 ACM on Asia Conference on Computer and Communications Security. 2022: 1098-1112.

[10] Huang Y, Evans D, Katz J. Private set intersection: Are garbled circuits better than custom protocols?[C]//NDSS. 2012.

[11] Yao A C C. How to generate and exchange secrets[C]//27th annual symposium on foundations of computer science (Sfcs 1986). IEEE, 1986: 162-167.

[12] Goldreich O, Micali S, Wigderson A. How to play any mental game, or a completeness theorem for protocols with honest majority[M]//Providing Sound Foundations for Cryptography: On the Work of Shafi Goldwasser and Silvio Micali. 2019: 307-328.

[13] Kolesnikov V, Schneider T. Improved garbled circuit: Free XOR gates and applications[C]//Automata, Languages and Programming: 35th International Colloquium, ICALP 2008, Reykjavik, Iceland, July 7-11, 2008, Proceedings, Part II 35. Springer Berlin Heidelberg, 2008: 486-498.

[14] Rosulek M, Roy L. Three halves make a whole? beating the half-gates lower bound for garbled circuits[C]//Annual International Cryptology Conference. Cham: Springer International Publishing, 2021: 94-124.

[15] Pinkas B, Schneider T, Weinert C, et al. Efficient circuit-based PSI via cuckoo hashing[C]//Annual International Conference on the Theory and Applications of Cryptographic Techniques. Cham: Springer International Publishing, 2018: 125-157.